%% file: main.tex
\documentclass[sigconf]{acmart}

\usepackage{float}
\usepackage{subcaption}
\usepackage{tabularx}
\usepackage{amsmath}

\usepackage{amssymb}
\usepackage{balance}
\usepackage{booktabs}
\usepackage{caption}
\usepackage{colortbl}
\usepackage{xcolor}
\usepackage{enumitem}
\usepackage{comment}
\usepackage{soul}

\newcommand{\answerbox}[1]{
\begin{center}
\fcolorbox{black}{gray!8}{
\begin{minipage}{0.95\columnwidth}
#1
\end{minipage}}
\end{center}
}

\title{Patterns of Bot Participation and Emotional Influence in Open-Source Development}

\author{Matteo Vaccargiu}
\affiliation{
  \institution{University of Cagliari}
  \city{Cagliari}
  \country{Italy}
}
\email{matteo.vaccargiu@unica.it}

\author{Riccardo Lai}
\affiliation{
  \institution{University of Cagliari}
  \city{Cagliari}
  \country{Italy}
}
\email{r.lai39@studenti.unica.it}

\author{Maria Ilaria Lunesu}
\affiliation{
  \institution{University of Cagliari}
  \city{Cagliari}
  \country{Italy}
}
\email{mariai.lunesu@unica.it}

\author{Andrea Pinna}
\affiliation{
  \institution{University of Cagliari}
  \city{Cagliari}
  \country{Italy}
}
\email{andrea.pinna83@unica.it}

\author{Giuseppe Destefanis}
\affiliation{
  \institution{University College London}
  \city{London}
  \country{United Kingdom}
}
\email{g.destefanis@ucl.ac.uk}

\begin{document}

\copyrightyear{2026}
\acmYear{2026}
\setcopyright{cc}
\setcctype{by}
\acmConference[BoatSE '26]{7th International Workshop on Bots and Agents in Software Engineering }{April 12--18, 2026}{Rio de Janeiro, Brazil}
\acmBooktitle{7th International Workshop on Bots and Agents in Software Engineering (BoatSE '26), April 12--18, 2026, Rio de Janeiro, Brazil}
\acmDOI{10.1145/3786161.3788455}
\acmISBN{979-8-4007-2393-3/2026/04}

\begin{abstract}
We study how bots contribute to open-source discussions in the Ethereum ecosystem and whether they influence developers' emotional tone. Our dataset covers 36,875 accounts across ten repositories with 105 validated bots (0.28\%). Human participation follows a U-shaped pattern, while bots engage in uniform (pull requests) or late-stage (issues) activity. Bots respond faster than humans in pull requests but play slower maintenance roles in issues. Using a model trained on 27 emotion categories, we find bots are more neutral, yet their interventions are followed by reduced neutrality in human comments, with shifts toward gratitude, admiration, and optimism and away from confusion. These findings indicate that even a small number of bots are associated with changes in both timing and emotional dynamics of developer communication.
\end{abstract}



\maketitle

\input{introduction}
\input{relatedworks}
\input{methodology}
\input{findings}
\input{threats}
\input{conclusion}

\balance
\bibliographystyle{ACM-Reference-Format}

\bibliography{biblio}

\end{document}

%% file: introduction.tex
\vspace{-6pt}
\section{Introduction}

Bots are now a common presence in open-source repositories. They manage dependencies, run automated tests, and enforce code-quality checks (e.g., via static-analysis tools) as part of CI/CD and code-review workflows \cite{wessel2020expect}. These bots routinely operate alongside human contributors in issues, pull requests, and commits, posting automated comments and pull requests during normal development. However, despite this prevalence, relatively little is known about bots’ conversational behavior or how their presence shapes human communication in these threads \cite{penzenstadler2022bots}.

Most prior work has assessed bots from a technical perspective – for example, evaluating correctness or efficiency of the automated tasks they perform \cite{santhanam2022bots}. Far fewer studies examine bots’ social or behavioral role in developer communities (e.g., whether their activity resembles that of human contributors over a discussion’s life cycle). In contrast, the social dimension of open-source collaboration is well established: developer communication involves emotions and affect that influence collaboration, trust, and productivity. Empirical studies show that the sentiment expressed in developer comments can significantly affect outcomes, including review efficiency, collaboration dynamics, and discussion focus in blockchain projects \cite{el2019empirical, 10.1145/3661167.3661194, VACCARGIU2026108003}. Introducing automated bot 
messages raises questions about whether they stabilize conversations, reduce 
emotional variability, or produce different shifts in tone. This study addresses these gaps through an analysis of multiple open-source repositories, combining temporal measurements of discussion lifecycles with an examination of emotional profiles of comments. The work is guided by the following research questions:

\textbf{RQ1 — How do bots participate in open-source discussions compared to human contributors?}
This question investigates whether bots mirror human interaction patterns in software repositories — in terms of activity timing and engagement dynamics. Specifically, do bots exhibit the same life-cycle ``U-shape'' observed in human discussions (early engagement, mid silence, and pre-closure surge), or do they follow distinct temporal patterns?

\textbf{RQ2 — Do bots influence the emotional tone of developer discussions?}
This question examines whether bot interventions alter the emotional profile of subsequent human communication. Using emotion distributions (27 labels from RoBERTa/GoEmotions), we measure pre- and post-bot differences to assess whether bots stabilize, neutralize, or amplify emotional variability in ongoing threads.

%% file: relatedworks.tex
\vspace{-9pt}
\section{Related Works}

Software Engineering Bots (or DevBots)~\cite{10.1145/2950290.2983989, 10.1145/3368089.3409680} entered the software development process to improve productivity, helping developers make decisions and facilitating communication. Storey and Zagalsky~\cite{10.1145/2950290.2983989} recognized that bots improve efficiency and effectiveness, performing repetitive tasks and improving decision-making and knowledge distribution. Erlenhov et al.~\cite{10.1145/3368089.3409680} illustrated three bot personas: chatbot persona (natural language interface), autonomous bot persona (independent tools triggered by project events), and smart bot persona (capable of complex tasks like bug fixing).

\vspace{-4pt}

\paragraph{Bot accounts in version control systems}
Wessel et al.~\cite{10.1145/3274451} found significant bot adoption in open-source projects. Wei et al.~\cite{11050809} examined smart bots in Apache pull requests, revealing substantial bot involvement in leaving comments. Lambiase et al.~\cite{10.1145/3704806} studied motivations behind bot adoption. Rebatchi et al.~\cite{rebatchi2024dependabot} measured Dependabot effectiveness in security-related pull requests, revealing high developer receptiveness. Choksi et al.~\cite{10.1145/3630106.3659019} discussed how tools like Dependabot and Copilot reinforce centralization tendencies in open-source development. Chidambaram et al. introduced a comprehensive dataset~\cite{chidambaram2023dataset} of bot and human activities on GitHub, proposed BIMBAS~\cite{CHIDAMBARAM2025112287} and RABBIT~\cite{chidambaram2024rabbit} for automated bot identification; we developed a custom approach combining rule-based heuristics with manual validation to prioritize precision for emotion analysis. While these studies examine bot adoption patterns, identification methods and aggregate project-level sentiment, our work investigates when bots participate during discussion lifecycles, whether their participation timing differs between artifact types and if the emotional tone shifts before and after bot interventions using fine-grained emotion classification across 27 categories.

\vspace{-4pt}

\paragraph{Effectiveness and emotional influence of bot comments in the development workflow}

Investigating bot impact on developer emotions and choices is essential~\cite{10172895}. Brown and Parnin~\cite{8823645} found low success rates for recommender bots, suggesting bots alone cannot influence developer behavior without social context knowledge. Farah et al.~\cite{10.1145/3528228.3528409} investigated reactions to bot comments on GitHub, finding laugh reactions used both humorously and in response to unexpected bot behavior. Wang et al.~\cite{wang2023more} showed that emoji reactions facilitate collaborative communication during reviews.

Gao et al.~\cite{gao2022does, 9930282} analyzed sentiment in developer comments, showing developers exhibit more neutral sentiment on projects using bots, with no significant differences after bot integration. Saadat et al.~\cite{9474411} examined workflow modifications, finding that in only-human teams comments cluster together, while in human-bot teams they scatter throughout event sequences, possibly due to reduced awareness of automation introduction.

Eshraghian~\cite{10.1108/ITP-01-2023-0084, eshraghian2023dynamics} investigated emotions around AI tool integration by analyzing tweets about GitHub Copilot, identifying six emotions: challenge, achievement, loss, deterrence, scepticism, and apathy. The study explored how these emotions evolved over time, shifting from negative to positive as developers focused on AI capabilities enhancing professional identity rather than as threats.

Our work differs from prior research by combining temporal lifecycle analysis with fine-grained emotion classification. While existing studies examine aggregate sentiment~\cite{gao2022does, 9930282} or bot adoption patterns~\cite{10.1145/3704806, 10.1145/3274451, 10.1145/3510003.3512765}, we investigate conversation-level emotional dynamics before and after individual bot interventions across 27 emotion categories, revealing artifact-specific participation patterns.

%% file: methodology.tex
\vspace{-5pt}
\section{Methodology}\label{sec:dataset}

\subsection{Dataset}
Our analysis examines ten repositories from the Ethereum ecosystem, building upon the dataset introduced by Vaccargiu et al.~\cite{11025661}, who conducted a longitudinal study of contributor dynamics across these same repositories spanning a decade of development. We extended this dataset by applying bot and emotion detection to all comments across the repositories.

\paragraph{Project Selection and Ecosystem}
We analyzed ten open-source projects from the Ethereum blockchain development ecosystem, selected to represent different layers and roles within a connected software infrastructure. The projects comprise: \textbf{Core Protocol Implementations} (go-ethereum, consensus-specs, solidity), \textbf{Developer Libraries} (ethers.js, web3.js), \textbf{Development Tools and Frameworks} (hardhat, truffle, metamask, openzeppelin), and \textbf{Oracle and Data Infrastructure} (chainlink).

This interconnected ecosystem is particularly suitable for bot research because: (1) projects share common infrastructure enabling consistent bot deployment patterns, (2) high project visibility and active communities ensure diverse bot types and sufficient data for statistical analysis, and (3) direct dependency relationships and shared maintainers allow studying bot participation across project types while controlling for domain-specific factors. All projects use GitHub for development coordination, enabling consistent data collection.

\paragraph{Bot Detection} To identify automated accounts in our dataset, we implemented a three-step bot detection framework.

The first step applies deterministic rules to identify bot accounts based on specific patterns. We implemented five detection criteria: (1) username pattern matching (e.g., bot\$, [bot], dependabot), (2) GitHub User Type: BOT metadata, (3) high activity with low diversity (top 5th percentile activity, Shannon entropy < 0.5), (4) high message repetitiveness (>95\%), and (5) pure commit bots (100\% commit ratio). Accounts satisfying any criterion were assigned \texttt{is\_bot\_rule = True} and received a base bot score of 60 points.

For accounts not flagged by Step 1 ($n = 36{,}548$), we applied Isolation Forest~\cite{liu2008isolation}, an unsupervised anomaly detection algorithm. We extracted 19 behavioral features including temporal patterns (inter-event times, activity entropy, circadian patterns), behavioral metrics (activity diversity, action type ratios), and content characteristics (message length, repetitiveness). Features were standardized using z-score normalization. The Isolation Forest was configured with contamination parameter $\alpha = 0.1$ and 100 estimators, identifying 3,655 anomalous accounts. Anomaly scores were normalized to contribute 0-30 points to the final bot score.

The final bot score (0-100) was calculated as:
\[ \text{bot\_score} = w_1 \cdot \mathbb{1}_{rule} + w_2 \cdot s_{iso} + w_3 \cdot r_{msg} + w_4 \cdot (1 - d_{act})\]
where $\mathbb{1}_{rule}$ indicates Step 1 detection ($w_1 = 60$), $s_{iso}$ is the normalized isolation score ($w_2 = 30$), $r_{msg}$ is message repetitiveness ($w_3 = 5$), and $d_{act}$ is normalized activity diversity ($w_4 = 5$). Accounts with $\text{bot\_score} \geq 50$ were classified as bots.

The automated system identified 337 bot accounts (0.91\% of 36,885 total accounts). Two authors independently examined each account, considering username patterns, profile information, activity patterns, and known bot services. The manual validation revealed 243 false positives (72.1\%), 11 false negatives, and 10 ghost accounts (excluded from analysis). After manual corrections, the final dataset contained 36,875 accounts, comprising 105 confirmed bots (0.28\%) and 36,770 human contributors (99.72\%). The low automated precision (27.9\%) reflects a conservative strategy prioritizing recall; all subsequent analyses use only the 105 manually validated bots, ensuring reliability. The corrected system achieved recall of 89.5\% and accuracy of 99.3\%.

Bot participation varies substantially across artifact types, ranging from 0.06\% to 56.90\% in comments, 0\% to 2.50\% in issues, 0\% to 3.66\% in commits, and 0\% to 13.59\% in pull requests. Several repositories show zero bot activity in specific artifact types. Our analysis does not distinguish between bot types (CI/CD, dependency management, security scanning), potentially obscuring type-specific patterns.

\vspace{-5pt}

\paragraph{Emotion Detection} For emotion detection, we employed the \texttt{roberta-base-go\_emotions}\footnote{\url{https://huggingface.co/SamLowe/roberta-base-go_emotions}} model, which uses the RoBERTa architecture to classify text into \textit{27 distinct emotions}. This model's ability to handle multi-label emotion detection made it particularly suitable for analyzing the complex emotional content present in developer communications. We followed the emotion analysis procedure established by Vaccargiu et al.~\cite{11024360}, applying the model to all comments in our dataset to generate probability distributions over the emotion categories. Each comment receives a probability vector where values sum to unity, representing the model's confidence in the presence of each emotion category. These probability vectors serve as the foundation for all subsequent emotional analyses.

\subsection{Analysis Methods}\label{sec:methodology}

Our analytical approach examines bot participation patterns and emotional 
dynamics through statistical methods applied to GitHub artifacts. We analyze 
temporal positioning of comments, measure response times, and quantify emotional 
divergence using probability distributions from pre-trained models. All analyses 
employ non-parametric tests with effect sizes reported alongside significance 
tests.

\subsubsection{Temporal Patterns of Bot Participation (RQ1)}

To investigate how bots participate in open-source discussions compared to human contributors, we analyzed comment timing patterns across two dimensions: positioning within discussion lifecycles and response latency following bot interventions.

\paragraph{Comment Position Within Thread Lifecycle}
For each discussion thread with valid temporal bounds (creation timestamp preceding closure timestamp), we normalized comment positions to a $[0,1]$ scale representing the thread lifecycle:
\begin{equation}
\text{position} = 
\frac{t_{\text{comment}} - t_{\text{thread\_created}}}
     {t_{\text{thread\_closed}} - t_{\text{thread\_created}}}
\end{equation}
where position 0 represents thread opening and position 1 represents closure. We aggregated normalized positions by user type (bot versus human) and compared the resulting distributions using the two-sample Kolmogorov--Smirnov test, which assesses whether two samples originate from the same underlying distribution without assuming parametric forms. Effect sizes were quantified through Cliff's Delta, with magnitudes interpreted as negligible, small, medium, or large according to established thresholds.

\paragraph{Response Time Analysis}
To examine interaction dynamics, we measured human response latency following bot comments. Within each thread, comments were ordered chronologically, and for every bot comment, we identified the first subsequent human comment in the same thread. Response time $\Delta t$ (in hours) was computed as:
\begin{equation}
\Delta t = 
\frac{t_{\text{first human after bot}} - t_{\text{bot}}}{3600}
\end{equation}
We summarized response time distributions per repository using median and mean statistics, and visualized the distributions through histograms and empirical cumulative distribution functions.

\subsubsection{Emotional Patterns Around Bot Interventions (RQ2)}

To assess whether bot participation influences the emotional tone of developer discussions, we conducted three complementary analyses: comparing baseline emotional profiles between bots and humans, measuring emotional shifts in human comments before versus after bot interventions, and testing differences in specific emotion categories.

\paragraph{Baseline Emotional Divergence}
Each comment in our dataset contains a probability vector over 27 emotion categories (derived from the RoBERTa-GoEmotions model), with probabilities summing to unity. For each repository and user type, we computed mean emotion probability distributions, yielding two normalized vectors representing the typical emotional profile of bot versus human communication. We quantified the divergence between these profiles using the Jensen--Shannon divergence (JSD) with base-2 logarithm, a symmetric and bounded measure of distributional difference ranging from 0 (identical distributions) to 1 (maximally different distributions). To assess whether observed divergence values significantly exceed chance expectations, we employed right-tailed permutation tests with 10,000 iterations. In each iteration, user type labels were randomly reassigned, mean emotion vectors were recomputed for the permuted groups, and JSD was calculated. The empirical p-value was estimated as the proportion of permuted divergence values meeting or exceeding the observed value.

\paragraph{Pre-Post Emotional Shifts Around Bot Comments}
To examine emotional patterns in human comments before and after bot interventions, we analyzed threads containing at least one bot comment with human comments both preceding and following the bot timestamp. For each qualifying thread, we partitioned human comments into pre-bot and post-bot sets based on their temporal relationship to the bot comment. Mean emotion vectors were computed for each partition, where $\overline{\mathbf{e}}_{\text{pre}}$ and $\overline{\mathbf{e}}_{\text{post}}$ represent the average emotion distributions before and after bot intervention. We calculated the Jensen-Shannon Divergence (JSD) between these distributions, $\operatorname{JSD}(\overline{\mathbf{e}}_{\text{pre}},\,\overline{\mathbf{e}}_{\text{post}})$, to quantify the magnitude of emotional shift associated with bot intervention. We tested whether the distribution of thread-level JSD values significantly exceeds zero using the one-sided Wilcoxon signed-rank test. To understand the directionality of emotional changes, we computed mean differences for individual emotion categories across all analyzed threads.

\paragraph{Single Emotion Comparison}
To test our specific hypothesis regarding emotional neutrality, we compared the probability of the neutral emotion category between bot and human comments using the Mann--Whitney U test with a one-sided alternative hypothesis (bot neutrality greater than human neutrality). Cliff's Delta was computed to quantify the effect size.
All statistical tests were conducted at $\alpha = 0.05$ significance level, with both p-values and effect size estimates reported to distinguish statistical significance from practical importance. Complete results, including intermediate calculations and test statistics, were preserved to ensure reproducibility of all analyses.

%% file: findings.tex
\vspace{-10pt}
\section{Findings}

\subsection{RQ1 — How do bots participate in open-source discussions compared to human contributors?}

To understand bot participation patterns in open-source development, we analyzed 36,875 contributor accounts across ten repositories, including core blockchain implementations (go-ethereum, consensus-specs), developer libraries (ethers.js, web3.js), and development frameworks (hardhat, truffle). After applying our three-stage detection framework and manual validation, we identified 105 bot accounts (0.28\%) and 36,770 human contributors (99.72\%).

\vspace{-5pt}

\paragraph{Temporal Patterns in Discussion Lifecycles}

To examine when bots participate during discussion lifecycles, we normalized all issues and pull requests to a 0-1 scale, where 0 represents opening and 1 represents closure. Human contributors exhibit a consistent U-shaped participation pattern across all repositories: high activity at opening, minimal activity during the middle phase, and renewed activity before closure. This pattern holds consistently across both issues and pull requests.

\begin{figure}[h]
\centering
\includegraphics[width=\columnwidth]{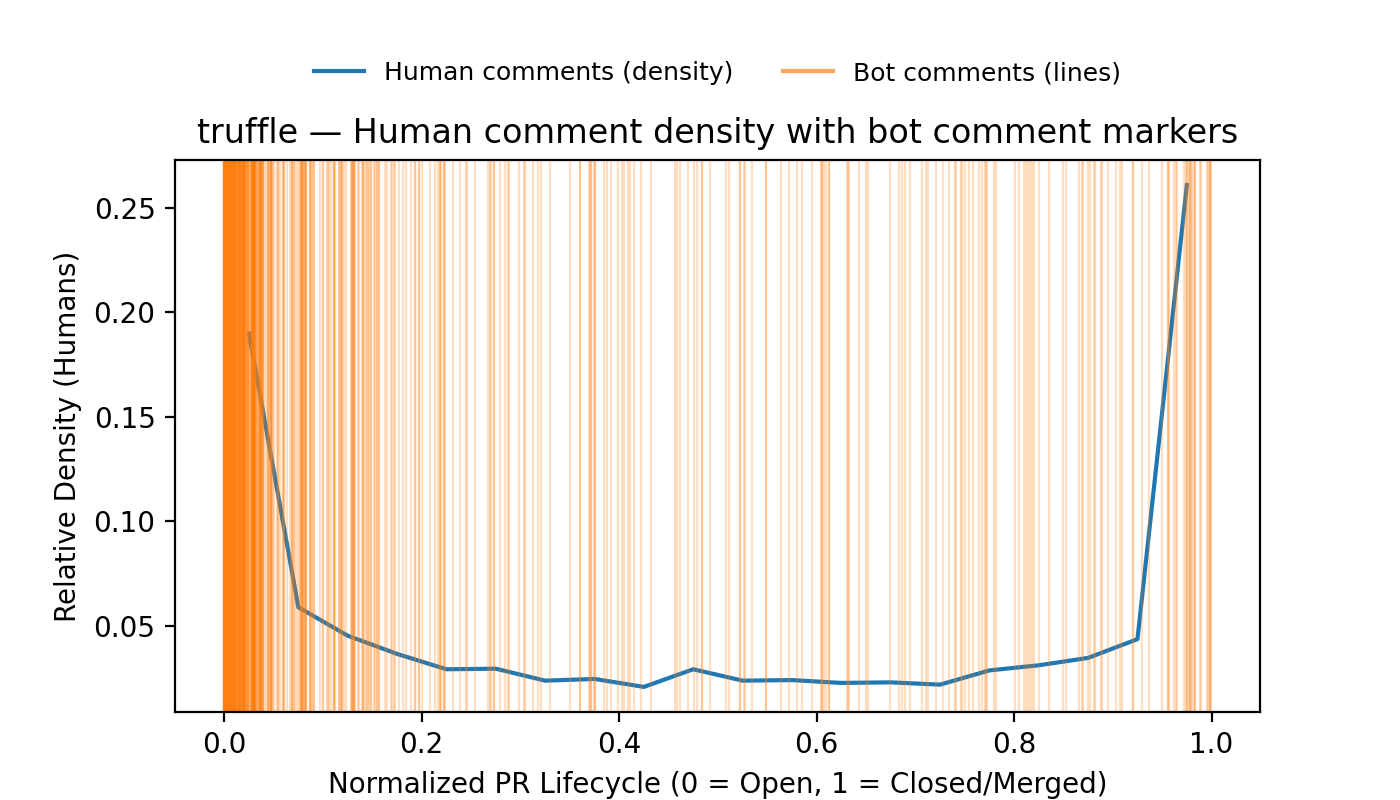}
\caption{Pull request lifecycle in truffle: human comment density (blue) shows 
U-shaped pattern; bot markers (orange) show early concentration with continuous 
activity.}
\label{fig:pr_lifecycle}
\end{figure}

Bot participation patterns, however, differ between artifact types. In pull request discussions, bots demonstrate continuous presence throughout the lifecycle. Repositories like solidity, web3.js, and go-ethereum show dense, uniform bot activity from opening to closure, creating a pattern of sustained automated engagement. Other repositories like truffle (Figure~\ref{fig:pr_lifecycle}) and openzeppelin show heavy bot clustering at pull request opening (normalized position 0.0), followed by continued activity throughout the lifecycle. The human U-shaped pattern remains visible in all cases, but bot activity does not follow this pattern, instead appearing relatively constant across lifecycle phases.

\vspace{-11pt}

\begin{figure}[h]
\centering
\includegraphics[width=\columnwidth]{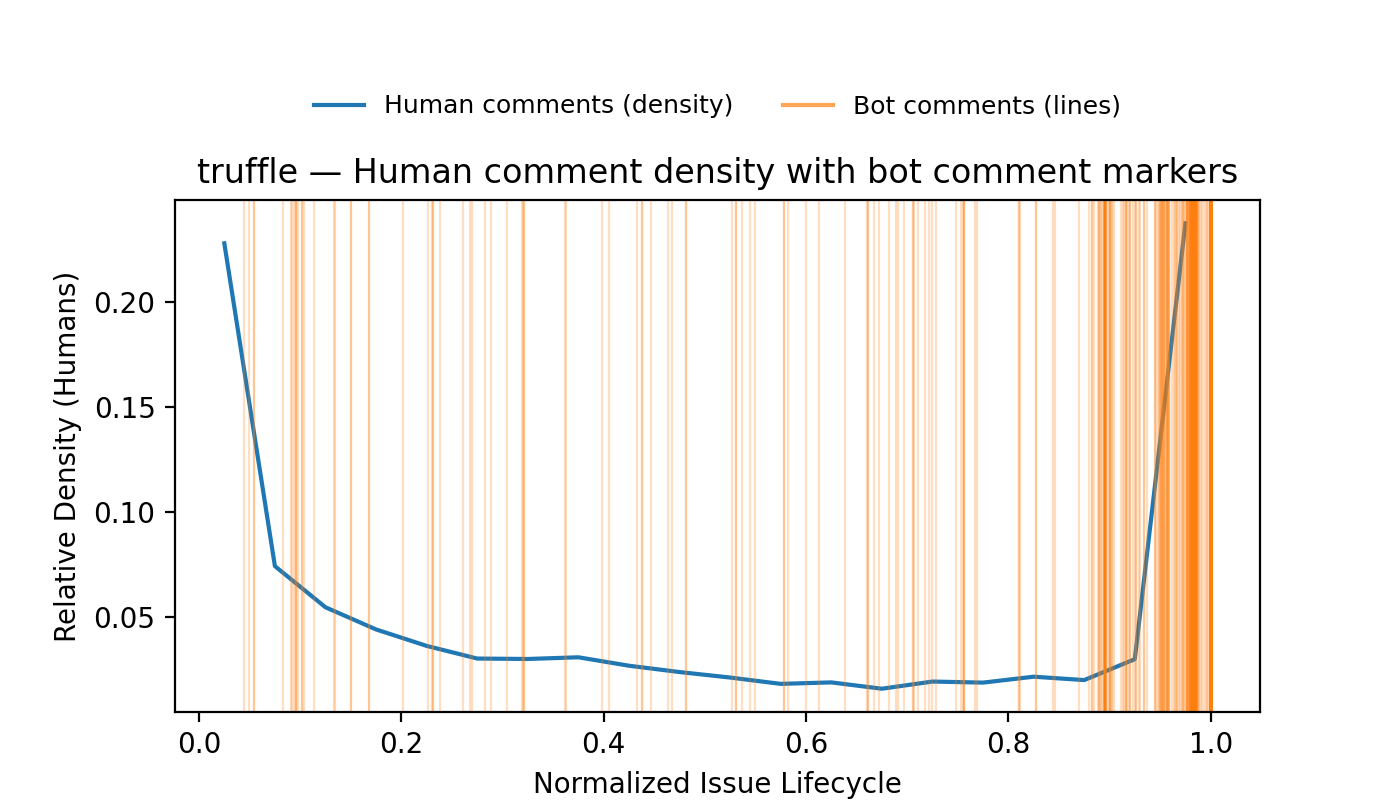}
\caption{Issue lifecycle in truffle: bot activity concentrates late-stage.}
\label{fig:issue_lifecycle}
\end{figure}

\vspace{-6pt}
Issue discussions present a contrasting pattern. While human contributors maintain their U-shaped engagement pattern, bot participation shows greater variability. Some repositories like metamask and solidity display relatively uniform bot distribution across the issue lifecycle. Others like go-ethereum and truffle (Figure~\ref{fig:issue_lifecycle}) show bot clustering toward later lifecycle stages (normalized position 0.8-1.0), suggesting automation focused on issue closure or staleness detection. Web3.js and chainlink demonstrate bot concentration at both endpoints (positions 0.0-0.1 and 0.9-1.0), partially mirroring the human pattern but with substantially different timing.
    
\vspace{-10pt}

\vspace{-10pt}
\paragraph{Response Time Analysis}

We analyzed first response times for 23,937 threads containing bot responses and 8,941 threads containing both bot and human responses. Overall, bots demonstrate substantially faster response times than humans, with a median of 2,075 seconds (34.6 minutes) compared to the human median of 58,497 seconds (16.2 hours). The response time distributions differ significantly (Kolmogorov-Smirnov test: $D = 0.347$, $p < 0.001$; Cliff's Delta $\delta = -0.384$, medium effect size), with the negative delta indicating that bots respond faster than humans.

However, this speed advantage varies dramatically by artifact type. In pull request discussions (23,490 threads), bots achieve a median first response time of 2,021 seconds (33.7 minutes), while humans require a median of 56,308 seconds (15.6 hours). In threads with both bot and human responses (8,617 threads), bots respond first in 70.1\% of cases. Response speeds vary substantially across repositories. Three repositories show near-instantaneous bot responses: go-ethereum (median 0 minutes, 96.5\% bot-first), hardhat (median 0.1 minutes, 97.5\% bot-first), and openzeppelin (median 0.1 minutes, 89.4\% bot-first). In contrast, chainlink shows considerably slower bot responses (median 687 minutes or 11.5 hours, with bots responding first in only 31.8\% of cases).

Issue discussions present an inverted pattern. Across 447 issue threads, bots show a median first response time of 5,202,537 seconds (approximately 60 days), while humans respond in a median of 343,891 seconds (4.0 days). In threads with both response types (324 threads), humans respond first in 75.6\% of cases. This reversal indicates that bots play a substantially different role in issue discussions compared to pull requests.

\paragraph{Statistical Validation of Lifecycle Patterns} Kolmogorov-Smirnov tests validate the observed temporal patterns. For issue comments (2,344 bot, 108,947 human), distributions differ significantly (($D = 0.345$, $p = 6.0 \times 10^{-245}$; Cliff's Delta $\delta = 0.304$). The positive delta indicates that bots comment significantly later in issue lifecycles than humans. For pull request comments (47,984 bot comments, 72,424 human comments), the distributions also differ significantly ($D = 0.160$, $p < 0.001$; Cliff's Delta $\delta = -0.218$). The negative delta indicates that bots comment significantly earlier in pull request lifecycles compared to humans. These tests confirm genuine differences in participation timing, not random variation.

\answerbox{
\textbf{Answer to RQ1}: Bots participate differently than human contributors. While representing less than 1\% of contributors, bots concentrate activity in discussion comments. In pull requests, bots respond rapidly (median 34 minutes) and maintain continuous activity throughout the lifecycle. In issues, bots respond substantially slower (median 60 days) and engage later in the lifecycle. These patterns suggest bot participation is optimized for different purposes: continuous integration and automated checks in pull requests versus maintenance tasks such as stale issue detection in issues.}

\subsection{RQ2 — Do bots influence the emotional tone of developer discussions?}

To examine whether bot participation influences the emotional characteristics of developer discussions, we applied a pre-trained emotion classification model (RoBERTa fine-tuned on GoEmotions) to all comments in our dataset. This model assigns probability distributions over 27 emotion categories plus a neutral category to each comment, with probabilities summing to 1.0. We analyzed 50,328 bot comments and 181,371 human comments across the ten repositories.
\vspace{-4pt}

\paragraph{Emotional Profiles of Bots and Humans}

\begin{figure}[h]
\centering
\includegraphics[width=\columnwidth]{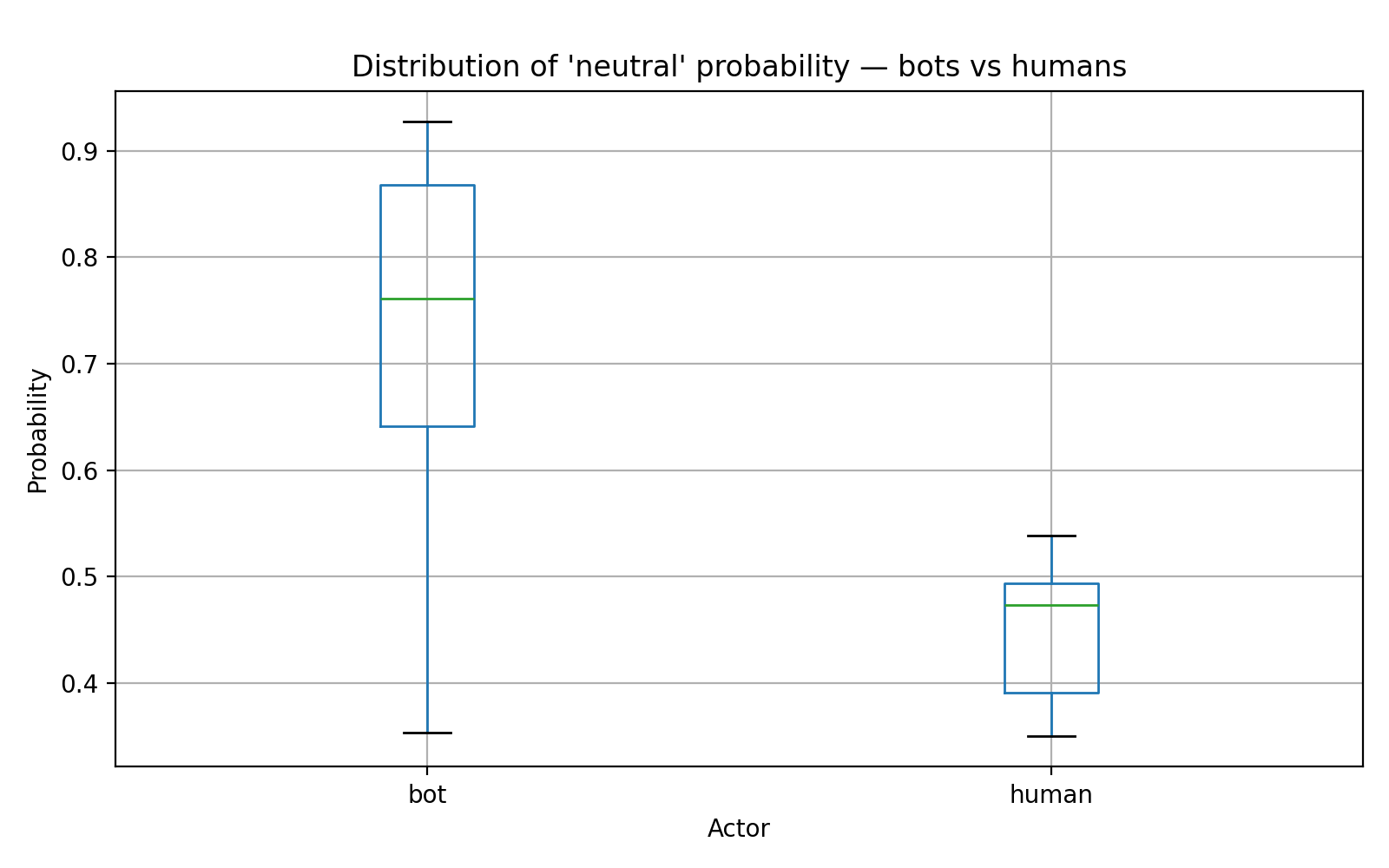}
\caption{Neutral emotion probability in bot vs. human comments across all repositories. Bots exhibit significantly higher neutral probability (median 0.76) compared to humans (median 0.47), with minimal overlap between distributions (Mann-Whitney U test: $p < 0.001$, Cliff's Delta $\delta = 0.740$).}
\label{fig:neutral_boxplot}
\end{figure}

We first examined the baseline emotional characteristics of bot and human comments by computing the average emotion probability distribution for each group within each repository. Across nine of eleven analyses (10 repos + 1 all aggregate) repositories, both bots and humans show neutral as the dominant emotion category. However, the degree of neutrality differs substantially between the two groups. Bot comments exhibit a median neutral probability of 0.76, while human comments show a median neutral probability of 0.47 (Figure~\ref{fig:neutral_boxplot}). This difference represents approximately 60\% higher neutrality in bot communications.

The distribution of neutral probability across repositories reveals consistent patterns. Bot neutral probabilities range from 0.35 to 0.93 across repositories, with seven repositories showing bot neutral probabilities exceeding 0.80. In contrast, human neutral probabilities cluster more tightly between 0.35 and 0.54. Statistical testing using Mann-Whitney U tests confirms that bots demonstrate significantly higher neutral probability than humans in nine of eleven cases (including the aggregate dataset), with all significant results showing $p < 0.001$. Seven repositories demonstrate large effect sizes (Cliff's Delta $> 0.5$), with web3.js, metamask, and hardhat showing the strongest effects at $\delta = 0.798$, $0.793$, and $0.769$ respectively. The overall effect across all repositories yields Cliff's Delta of 0.740, indicating a large and consistent difference.

One notable exception emerges in the core blockchain implementation go-ethereum, where bots show Cliff's Delta of $-0.224$ (not statistically significant). In this repository, bot comments express substantially more gratitude than other repositories, with an average gratitude probability of 0.57 compared to a neutral probability of 0.35. This inverted pattern reflects a deliberate design choice where bots acknowledge contributor efforts, contrasting with the purely informational bots prevalent in other repositories.

To quantify the overall divergence between bot and human emotional profiles, we calculated JSD between the emotion probability distributions for each repository. The JSD values range from 0.084 (openzeppelin) to 0.285 (ethers.js), indicating varying degrees of emotional difference across projects. Development frameworks and tools (ethers.js: 0.285, go-ethereum: 0.261, metamask: 0.192, web3.js: 0.188, hardhat: 0.178) show larger divergences, while smart contract development libraries (openzeppelin: 0.084, consensus-specs: 0.106, solidity: 0.126) show smaller differences.

We validated whether these observed differences exceed random variation using permutation tests. For each repository, we randomly shuffled bot and human labels 10,000 times and recalculated JSD under the null hypothesis that emotional profiles are equivalent. The observed JSD significantly exceeds the permutation distribution in nine of ten repositories ($p < 0.001$), with only consensus-specs showing no significant difference ($p = 0.346$). This confirms that the emotional divergence between bots and humans reflects genuine behavioral differences rather than sampling variation.

\paragraph{Emotional Shifts in Human Comments Following Bot Interventions} To assess whether bot comments influence subsequent human emotional expression, we analyzed 1,511 discussion threads containing both bot interventions and all human comments before and after bot participation. For each thread, we computed the average emotion probability distribution from all human comments preceding the first bot comment (pre-bot profile) and all human comments following the last bot comment (post-bot profile). We then measured the emotional shift using JSD between these pre- and post-bot distributions.

The distribution of pre-post JSD values shows substantial emotional changes following bot interventions. The median JSD of 0.224 indicates moderate emotional shifts, with 75\% of threads showing JSD values exceeding 0.10. We categorized threads by change magnitude: 377 threads (25.0\%) show minimal change (JSD $< 0.1$), 584 threads (38.6\%) show moderate change (0.1 $\leq$ JSD $< 0.3$), and 550 threads (36.4\%) show substantial change (JSD $\geq 0.3$). Repository-level analysis reveals variation in emotional change magnitude, with hardhat showing the highest median pre-post JSD of 0.378, followed by web3.js (0.297), openzeppelin (0.292), and truffle (0.255). Chainlink shows the lowest median change at 0.119.

Wilcoxon signed-rank tests assess whether the observed emotional changes significantly exceed zero. Nine of eleven analyses (including the aggregate dataset) demonstrate statistically significant emotional shifts ($p < 0.001$), with only consensus-specs and ethers.js showing non-significant results due to small sample sizes ($n = 1$ thread each). These results provide strong evidence that bot interventions systematically alter the emotional characteristics of subsequent human discussion.

To understand the direction of emotional changes, we examined shifts in individual emotion categories averaged across all 1,511 threads. The analysis reveals that human emotional expression is more varied after bot interventions, not more neutral. The neutral emotion probability decreases by 0.0113 (from 0.5545 to 0.5432), representing a 2.0\% reduction. This decrease in neutrality coincides with increases in nine specific emotions and decreases in eight others. The strongest increase occurs in gratitude, which rises by 0.0241 (42.1\% increase). Other emotions showing notable increases include optimism ($+0.0040$, 21.6\% increase), admiration ($+0.0043$, 17.4\% increase), and joy ($+0.0014$, 32.7\% increase). Conversely, confusion shows the largest decrease at $-0.0127$ (19.2\% reduction), followed by approval ($-0.0121$, 13.9\% reduction) and curiosity ($-0.0052$, 8.3\% reduction). These patterns suggest that bot interventions are associated with reduced uncertainty-related emotions and increased appreciation-related emotions, consistent with bots providing automated status information, test results, or procedural updates.

\textit{Temporal Context of Emotional Changes} To contextualize the timeframe within which emotional influence occurs, we measured response times from bot comments to subsequent human comments in the same thread. Analysis of these response times across repositories reveals substantial variation, with median response times ranging from 1.06 hours (ethers.js) to 29.60 hours (web3.js). Seven repositories show median response times below 24 hours, suggesting that the majority of human reactions to bot interventions occur within approximately one day. Examination of cumulative distribution functions indicates that between 50\% and 100\% of human responses occur within 24 hours across different repositories.

\answerbox{
\textbf{Answer to RQ2}: Bot participation is associated with changes in the emotional tone of developer discussions. Despite bots being 60\% more neutral than humans, human comments following bot interventions do not show increased neutrality. Instead, human comments following bot participation show decreased neutrality with specific emotional shifts: confusion-related emotions decrease (confusion -19.2\%, curiosity -8.3\%) while appreciation-related emotions increase (gratitude +42.1\%, admiration +17.4\%). These changes are statistically significant across multiple validation approaches (Wilcoxon, permutation, and Mann-Whitney U tests, all $p < 0.001$), with emotional shifts observed within approximately 24 hours.}

%% file: threats.tex
\section{Threats to Validity}\label{sec:threats}
We identify and discuss potential threats to the validity of our findings.\\
\noindent
\textit{Construct Validity:} Our three-step bot detection achieved 27.9\% precision and 89.5\% recall after manual validation by two researchers, yielding 105 confirmed bots. While high recall captured most bots, 11 false negatives suggest some remain undetected, and the small bot proportion (0.28\%) means misclassifications could impact findings. For emotion classification, we employed RoBERTa-GoEmotions, validated in prior software engineering research~\cite{11024360}. Our aggregation of probabilities into mean vectors may smooth individual variations, addressed by analyzing both aggregate and thread-level patterns. We normalize thread lifecycles to [0,1] scale, which may obscure differences between very short and very long threads, and treat all bot comments equally regardless of type, potentially obscuring bot-specific effects.\\
\noindent
\textit{Internal Validity:} We analyzed 1,511 threads containing bot comments with sufficient human comments before and after, which may introduce selection bias as threads attracting bot attention may differ systematically. Confounding factors including repository practices, team composition, and issue complexity could influence patterns beyond bot interventions. While our pre-post design controls for some confounds, correlation does not imply causation—emotional tone might evolve naturally as discussions progress.\\
\noindent
\textit{External Validity:} Our findings are based on ten Ethereum repositories with unique characteristics (market exposure, decentralized governance, transparency culture). While spanning different functional roles, all operate within the same ecosystem, and patterns may not generalize to traditional open-source or other blockchain projects. Our dataset spans repository inception through 2024 but does not model temporal evolution in bot capabilities or systematically distinguish between bot types (CI/CD, dependency management, security scanning). Repository-specific patterns (e.g., go-ethereum's gratitude-expressing bots) suggest heterogeneity that aggregate analysis may obscure.\\
\noindent
\textit{Conclusion Validity:} We employed non-parametric tests (Kolmogorov Smirnov, Wilcoxon signed-rank, Mann-Whitney U) with effect sizes (Cliff's Delta, r-values) reported alongside p-values. However, multiple tests increase Type I error possibility, and the small bot proportion (0.28\%) means even large effect sizes may represent subtle changes. Our emotional analysis aggregates distributions using JSD, which may obscure individual comment variations. Response time analysis captures only first human comments, potentially influenced by time zones and availability. Our data collection relies on GitHub's API, which may not capture all content changes and excludes private communications. However, GitHub's API is standard for repository research and provides reliable access to public development artifacts.

%% file: conclusion.tex
\section{Conclusion and Future Works}\label{sec:conclusion}
This study examined bot participation patterns and emotional dynamics in open-source discussions across ten Ethereum repositories, analyzing 105 validated bots among 36,875 accounts.
Our analysis revealed distinct participation patterns between bots and human contributors. In pull request discussions, bots respond rapidly and maintain relatively constant activity throughout the lifecycle, while in issue discussions they respond slowly and concentrate on late-stage maintenance tasks. These patterns suggest bots serve different roles depending on artifact type: continuous integration checks in pull requests versus staleness detection in issues. 

Regarding emotional dynamics, bot comments exhibit significantly higher neutrality than human comments. However, human comments following bot interventions show decreased neutrality, with increases in appreciation-related emotions (gratitude, admiration, optimism) and decreases in uncertainty-related emotions (confusion, curiosity).

For open source projects considering bot adoption, our findings suggest three deployment strategies: prioritizing rapid response bots for pull request workflows, using maintenance focused bots for late stage issue management, and designing bot messages that reduce uncertainty through clarification. The go-ethereum case shows that design choices such as expressing gratitude can influence emotional dynamics, indicating that bot tone and messaging matter alongside technical functionality. These findings are based on the Ethereum ecosystem, whose decentralized governance, market exposure, and visibility may limit generalization.

Future work could distinguish among bot types such as CI/CD, dependency management, and security scanning, and compare software ecosystems to assess whether these effects extend beyond blockchain-based projects.

%% file: biblio.bib
@inproceedings{liu2008isolation,
  title={Isolation forest},
  author={Liu, Fei Tony and Ting, Kai Ming and Zhou, Zhi-Hua},
  booktitle={2008 eighth ieee international conference on data mining},
  pages={413--422},
  year={2008},
  organization={IEEE}
}

@inproceedings{10.1145/2950290.2983989,
author = {Storey, Margaret-Anne and Zagalsky, Alexey},
title = {Disrupting developer productivity one bot at a time},
year = {2016},
isbn = {9781450342186},
publisher = {Association for Computing Machinery},
address = {New York, NY, USA},
url = {https://doi.org/10.1145/2950290.2983989},
doi = {10.1145/2950290.2983989},
booktitle = {Proceedings of the 2016 24th ACM SIGSOFT International Symposium on Foundations of Software Engineering},
pages = {928–931},
numpages = {4},
location = {Seattle, WA, USA},
series = {FSE 2016}
}

@INPROCEEDINGS{8823645,
  author={Brown, Chris and Parnin, Chris},
  booktitle={2019 IEEE/ACM 1st International Workshop on Bots in Software Engineering (BotSE)}, 
  title={Sorry to Bother You: Designing Bots for Effective Recommendations}, 
  year={2019},
  volume={},
  number={},
  pages={54-58},
  doi={10.1109/BotSE.2019.00021}}

@inproceedings{10.1145/3368089.3409680,
author = {Erlenhov, Linda and Neto, Francisco Gomes de Oliveira and Leitner, Philipp},
title = {An empirical study of bots in software development: characteristics and challenges from a practitioner’s perspective},
year = {2020},
isbn = {9781450370431},
publisher = {Association for Computing Machinery},
address = {New York, NY, USA},
url = {https://doi.org/10.1145/3368089.3409680},
doi = {10.1145/3368089.3409680},
booktitle = {Proceedings of the 28th ACM Joint Meeting on European Software Engineering Conference and Symposium on the Foundations of Software Engineering},
pages = {445–455},
numpages = {11},
location = {Virtual Event, USA},
series = {ESEC/FSE 2020}
}

@inproceedings{chidambaram2023dataset,
  title={A dataset of bot and human activities in GitHub},
  author={Chidambaram, Natarajan and Decan, Alexandre and Mens, Tom},
  booktitle={2023 IEEE/ACM 20th International Conference on Mining Software Repositories (MSR)},
  pages={465--469},
  year={2023},
  organization={IEEE}
}

@inproceedings{10.1145/3528228.3528409,
author = {Farah, Juan Carlos and Spaenlehauer, Basile and Lu, Xinyang and Ingram, Sandy and Gillet, Denis},
title = {An exploratory study of reactions to bot comments on GitHub},
year = {2022},
isbn = {9781450393331},
publisher = {Association for Computing Machinery},
address = {New York, NY, USA},
url = {https://doi.org/10.1145/3528228.3528409},
doi = {10.1145/3528228.3528409},
booktitle = {Proceedings of the Fourth International Workshop on Bots in Software Engineering},
pages = {18–22},
numpages = {5},
location = {Pittsburgh, Pennsylvania},
series = {BotSE '22}
}

@INPROCEEDINGS{9474411,
  author={Saadat, Samaneh and Colmenares, Natalia and Sukthankar, Gita},
  booktitle={2021 IEEE/ACM Third International Workshop on Bots in Software Engineering (BotSE)}, 
  title={Do Bots Modify the Workflow of GitHub Teams?}, 
  year={2021},
  volume={},
  number={},
  pages={1-5},
  doi={10.1109/BotSE52550.2021.00008}}

@article{10.1108/ITP-01-2023-0084,
    author = {Eshraghian, Farjam and Hafezieh, Najmeh and Farivar, Farveh and de Cesare, Sergio},
    title = {AI in software programming: understanding emotional responses to GitHub Copilot},
    journal = {Information Technology \& People},
    volume = {38},
    number = {4},
    pages = {1659-1685},
    year = {2024},
    month = {04},
    issn = {0959-3845},
    doi = {10.1108/ITP-01-2023-0084},
    url = {https://doi.org/10.1108/ITP-01-2023-0084},
    eprint = {https://www.emerald.com/itp/article-pdf/38/4/1659/9846400/itp-01-2023-0084.pdf},
}

@article{10.1145/3704806,
author = {Lambiase, Stefano and Catolino, Gemma and Palomba, Fabio and Ferrucci, Filomena},
title = {Motivations, Challenges, Best Practices, and Benefits for Bots and Conversational Agents in Software Engineering: A Multivocal Literature Review},
year = {2024},
issue_date = {April 2025},
publisher = {Association for Computing Machinery},
address = {New York, NY, USA},
volume = {57},
number = {4},
issn = {0360-0300},
url = {https://doi.org/10.1145/3704806},
doi = {10.1145/3704806},
journal = {ACM Computing Surveys},
month = dec,
articleno = {93},
numpages = {37}
}

@INPROCEEDINGS{11024360,
  author={Vaccargiu, Matteo and Neykova, Rumyana and Novielli, Nicole and Ortu, Marco and Destefanis, Giuseppe},
  booktitle={2025 IEEE/ACM 18th International Conference on Cooperative and Human Aspects of Software Engineering (CHASE)}, 
  title={More Than Code: Technical and Emotional Dynamics in Solidity's Development}, 
  year={2025},
  volume={},
  number={},
  pages={260-271},
  doi={10.1109/CHASE66643.2025.00036}}

@INPROCEEDINGS{11025661,
  author={Vaccargiu, M. and Aufiero, S. and Ba, C. and Bartolucci, S. and Clegg, R. and Graziotin, D. and Neykova, R. and Tonelli, R. and Destefanis, G.},
  booktitle={2025 IEEE/ACM 22nd International Conference on Mining Software Repositories (MSR)}, 
  title={Mining a Decade of Event Impacts on Contributor Dynamics in Ethereum: A Longitudinal Study}, 
  year={2025},
  volume={},
  number={},
  pages={552-563},
  doi={10.1109/MSR66628.2025.00088}}

@inproceedings{10.1145/3630106.3659019,
author = {Choksi, Madiha Zahrah and Mandel, Ilan and Widder, David and Shvartzshnaider, Yan},
title = {The Emerging Artifacts of Centralized Open-Code},
year = {2024},
isbn = {9798400704505},
publisher = {Association for Computing Machinery},
address = {New York, NY, USA},
url = {https://doi.org/10.1145/3630106.3659019},
doi = {10.1145/3630106.3659019},
booktitle = {Proceedings of the 2024 ACM Conference on Fairness, Accountability, and Transparency},
pages = {1971–1983},
numpages = {13},
location = {Rio de Janeiro, Brazil},
series = {FAccT '24}
}

@article{10.1145/3274451,
author = {Wessel, Mairieli and de Souza, Bruno Mendes and Steinmacher, Igor and Wiese, Igor S. and Polato, Ivanilton and Chaves, Ana Paula and Gerosa, Marco A.},
title = {The Power of Bots: Characterizing and Understanding Bots in OSS Projects},
year = {2018},
issue_date = {November 2018},
publisher = {Association for Computing Machinery},
address = {New York, NY, USA},
volume = {2},
number = {CSCW},
url = {https://doi.org/10.1145/3274451},
doi = {10.1145/3274451},
journal = {Proc. ACM Hum.-Comput. Interact.},
month = nov,
articleno = {182},
numpages = {19}
}

@article{CHIDAMBARAM2025112287,
title = {A bot identification model and tool based on GitHub activity sequences},
journal = {Journal of Systems and Software},
volume = {221},
pages = {112287},
year = {2025},
issn = {0164-1212},
doi = {https://doi.org/10.1016/j.jss.2024.112287},
url = {https://www.sciencedirect.com/science/article/pii/S0164121224003315},
author = {Natarajan Chidambaram and Alexandre Decan and Tom Mens}
}

@INPROCEEDINGS{11050809,
  author={Wei, Chenhao and Xiao, Lu and Liao, Ting and Zhao, Yutong},
  booktitle={2025 IEEE/ACM International Workshop on Bots in Software Engineering (BotSE)}, 
  title={The Secret Life of Bots in Pull Requests: An Empirical Study based on Apache Projects}, 
  year={2025},
  volume={},
  number={},
  pages={33-37},
  doi={10.1109/BotSE67031.2025.00015},
  ISSN={},
  month={April},}

@article{rebatchi2024dependabot,
  title={Dependabot and security pull requests: large empirical study},
  author={Rebatchi, Hocine and Bissyand{\'e}, T{\'e}gawend{\'e} F and Moha, Naouel},
  journal={Empirical Software Engineering},
  volume={29},
  number={5},
  pages={128},
  year={2024},
  publisher={Springer}
}

@inproceedings{eshraghian2023dynamics,
  title={Dynamics of emotions towards ai-powered technologies: A study of github copilot},
  author={Eshraghian, F and Hafezieh, N and Farivar, F and De Cesare, S},
  booktitle={Academy of Management Annual Meeting Proceedings},
  year={2023},
  organization={Academy of Management}
}

@inproceedings{gao2022does,
  title={How does bot affect developer’s sentiment: An empirical study on github issues and prs},
  author={Gao, Anze and Zhang, Yang and Wang, Tao and Chen, Sihao and Deng, Jinsheng},
  booktitle={2022 IEEE Smartworld, Ubiquitous Intelligence \& Computing, Scalable Computing \& Communications, Digital Twin, Privacy Computing, Metaverse, Autonomous \& Trusted Vehicles (SmartWorld/UIC/ScalCom/DigitalTwin/PriComp/Meta)},
  pages={1856--1861},
  year={2022},
  organization={IEEE}
}

@INPROCEEDINGS{9930282,
  author={Gao, Anze and Chen, Sihao and Wang, Tao and Deng, Jinsheng},
  booktitle={2022 IEEE 13th International Conference on Software Engineering and Service Science (ICSESS)}, 
  title={Understanding the Impact of Bots on Developers Sentiment and Project Progress}, 
  year={2022},
  volume={},
  number={},
  pages={93-96},
  doi={10.1109/ICSESS54813.2022.9930282},
  ISSN={2327-0594},
  month={Oct},}

@article{wang2023more,
  title={More than react: investigating the role of emoji reaction in github pull requests},
  author={Wang, Dong and Xiao, Tao and Son, Teyon and Kula, Raula Gaikovina and Ishio, Takashi and Kamei, Yasutaka and Matsumoto, Kenichi},
  journal={Empirical Software Engineering},
  volume={28},
  number={5},
  pages={123},
  year={2023},
  publisher={Springer}
}

@INPROCEEDINGS{10172895,
  author={Ghorbani, Amir and Cassee, Nathan and Robinson, Derek and Alami, Adam and Ernst, Neil A. and Serebrenik, Alexander and Wąsowski, Andrzej},
  booktitle={2023 IEEE/ACM 45th International Conference on Software Engineering (ICSE)}, 
  title={Autonomy Is An Acquired Taste: Exploring Developer Preferences for GitHub Bots}, 
  year={2023},
  volume={},
  number={},
  pages={1405-1417},
  doi={10.1109/ICSE48619.2023.00123},
  ISSN={1558-1225},
  month={May},}

@inproceedings{wessel2020expect,
  title={What to expect from code review bots on GitHub? a survey with OSS maintainers},
  author={Wessel, Mairieli and Serebrenik, Alexander and Wiese, Igor and Steinmacher, Igor and Gerosa, Marco A},
  booktitle={Proceedings of the XXXIV Brazilian Symposium on Software Engineering},
  pages={457--462},
  year={2020}
}

@ARTICLE{penzenstadler2022bots,
  author={Penzenstadler, Birgit and Abrahão, Silvia and Staron, Miroslaw and Serebrenik, Alexander and Carver, Jeffrey C. and Hochstein, Lorin},
  journal={IEEE Software}, 
  title={Bots in Software Engineering}, 
  year={2022},
  volume={39},
  number={5},
  pages={101-104},
  doi={10.1109/MS.2022.3180906}}

@article{santhanam2022bots,
  title={Bots in software engineering: a systematic mapping study},
  author={Santhanam, Sivasurya and Hecking, Tobias and Schreiber, Andreas and Wagner, Stefan},
  journal={PeerJ Computer Science},
  volume={8},
  pages={e866},
  year={2022},
  publisher={PeerJ Inc.}
}

@article{el2019empirical,
  title={An empirical study of sentiments in code reviews},
  author={El Asri, Ikram and Kerzazi, Noureddine and Uddin, Gias and Khomh, Foutse and Idrissi, MA Janati},
  journal={Information and Software Technology},
  volume={114},
  pages={37--54},
  year={2019},
  publisher={Elsevier}
}

@inproceedings{chidambaram2024rabbit,
  title={RABBIT: A tool for identifying bot accounts based on their recent GitHub event history},
  author={Chidambaram, Natarajan and Mens, Tom and Decan, Alexandre},
  booktitle={Proceedings of the 21st International Conference on Mining Software Repositories},
  pages={687--691},
  year={2024}
}

@inproceedings{10.1145/3510003.3512765,
author = {Wessel, Mairieli and Abdellatif, Ahmad and Wiese, Igor and Conte, Tayana and Shihab, Emad and Gerosa, Marco A. and Steinmacher, Igor},
title = {Bots for pull requests: the good, the bad, and the promising},
year = {2022},
isbn = {9781450392211},
publisher = {Association for Computing Machinery},
address = {New York, NY, USA},
url = {https://doi.org/10.1145/3510003.3512765},
doi = {10.1145/3510003.3512765},
booktitle = {Proceedings of the 44th International Conference on Software Engineering},
pages = {274–286},
numpages = {13},
location = {Pittsburgh, Pennsylvania},
series = {ICSE '22}
}

@inproceedings{10.1145/3661167.3661194,
author = {Vaccargiu, Matteo and Aufiero, Sabrina and Bartolucci, Silvia and Neykova, Rumyana and Tonelli, Roberto and Destefanis, Giuseppe},
title = {Sustainability in Blockchain Development: A BERT-Based Analysis of Ethereum Developer Discussions},
year = {2024},
isbn = {9798400717017},
publisher = {Association for Computing Machinery},
address = {New York, NY, USA},
url = {https://doi.org/10.1145/3661167.3661194},
doi = {10.1145/3661167.3661194},
booktitle = {Proceedings of the 28th International Conference on Evaluation and Assessment in Software Engineering},
pages = {381–386},
numpages = {6},
location = {Salerno, Italy},
series = {EASE '24}
}

@article{VACCARGIU2026108003,
title = {Emotional expression in open- source: How project function shapes communication},
journal = {Information and Software Technology},
volume = {191},
pages = {108003},
year = {2026},
issn = {0950-5849},
doi = {https://doi.org/10.1016/j.infsof.2025.108003},
url = {https://www.sciencedirect.com/science/article/pii/S0950584925003428},
author = {Matteo Vaccargiu and Silvia Bartolucci and Nicole Novielli and Marco Ortu and Roberto Tonelli and Giuseppe Destefanis}
}
